\documentclass[preprint,12pt]{elsarticle}
\usepackage{bm}
\usepackage{amsmath}
\usepackage{amssymb}
\usepackage{xcolor,soul}
\usepackage{siunitx}
\newcommand{\correct}[1]{{#1}}

\journal{Scripta Materialia}

\begin{document}

\begin{frontmatter}

\title{Room-temperature vacancy emission from jog on edge dislocation in FCC nickel under glide force}

\author[1,2]{Yifan Wang\fnref{equal}}

\author[1]{Wu-Rong Jian\fnref{equal}}

\author[1]{Wei Cai\corref{corr-author}}
\ead{caiwei@stanford.edu}

\address[1]{Department of Mechanical Engineering, Stanford University, Stanford CA, 94305, USA}
\address[2]{Department of Materials Science and Engineering, Stanford University, Stanford CA, 94305, USA}

\fntext[equal]{Both authors contributed equally to the paper}
\cortext[corr-author]{Corresponding author}

\date{\today}

\begin{abstract}
Jogs, atomic-scale steps on dislocations, play an important role in crystal plasticity, yet they are often ignored in discrete dislocation dynamics (DDD) simulations due to their small sizes. While jogs on screw dislocations are known to move non-conservatively (i.e. climb) accompanied by vacancy emission, jogs on edge dislocations are commonly expected to move conservatively (i.e. glide) under ambient conditions. Here we report unexpected findings from molecular dynamics simulations of an edge dislocation containing a pair of unit jogs in face-centered cubic nickel at 300K. We observe that one of the jogs climbs and emits vacancies intermittently at higher stresses, unexpected at such a low temperature, as climb is typically associated with creep at roughly half of the melting temperature. Our results highlight the significance of the complex interplay between point defects (i.e., vacancies) and dislocations in room-temperature plasticity, suggesting that these interactions may be more significant than previously thought.
\end{abstract}


\end{frontmatter}


Dislocations are line defects in crystalline solids, and dislocation glide on slip planes is the primary mechanism of plastic deformation under ambient conditions~\cite{cai_imperfections_2016}.
When dislocations on intersecting slip planes move past each other, they may leave atomic-scale steps on themselves called jogs, where their slip plane changes to adjacent parallel planes~\cite{hull_introduction_2011}.
%
%
%
%
%
Jogs participate in various deformation mechanisms such as 
jog dragging~\cite{messerschmidt_model_1970}, climb~\cite{abu-odeh_insights_2020}, and cross-slip~\cite{vegge_atomistic_2001},
impacting macroscopic crystal plasticity behaviors such as creep~\cite{barrett_model_1965}, strain hardening~\cite{zbib_3d_2000}, and irradiation-mediated plasticity~\cite{rodney_dislocation_1999}.
However, it is challenging to incorporate the jog effects on dislocations into large-scale discrete dislocation dynamics (DDD) simulations of strain hardening under room temperature,
because of their small sizes and uncertainties about their atomistic mechanisms~\cite{bertin_frontiers_2020}.
%
%
%
Recent efforts have focused on incorporating jog effects into DDD models based on a set of assumptions~\cite{po_model_2022, li_coupled_2023}.
For example, jog motion together with a screw dislocation is out of its local glide plane and hence must be non-conservative (i.e., climb) accompanied by emission of point defects such as vacancies~\cite{cai_chapter_2004}.
Therefore, it has been proposed that the jogs on screw dislocations be modeled explicitly and assumed to be either completely pinned or to be movable with reduced mobility enabled by thermally activated climb~\cite{barrett_model_1965, gu_three-dimensional_2015, breidi_dislocation_2022}.
On the other hand, jogs on edge dislocations are geometrically allowed to move with the dislocation in a conservative manner requiring no climb~\cite{hirth_glide_1967},
although the conservative motion of the jog can still produce a drag on the edge dislocation due to a higher Peierls barrier for jog motion~\cite{strunk_jog_1975}.
In comparison, non-conservative jog motion (climb) along the edge dislocation is considered unlikely to occur at low temperatures~\cite{seeger_cxxxii_1955}.
%
%
%
As a result, no special treatment has been devised for jogs on edge dislocations in the DDD simulations framework.

In a recent work comparing DDD simulations with experimental compression tests of face-centered cubic (FCC) copper~\cite{akhondzadeh_direct_2023},
it was proposed that jogs are responsible for a significant reduction of the effective mobility of dislocations.
%
%
This hypothesis motivates a detailed study of the jog effect on dislocation mobility in FCC metals.
%
%
%
%
%
%
Early theoretical analysis of jog effects was based on the simple line-tension model~\cite{messerschmidt_model_1970, messerschmidt_model_1971}.
%
%
But this model does not capture the complexity of dislocations and jogs in FCC crystals, which often dissociate into partial dislocations~\cite{carter_extension_1979}.
%
Hirsch~\cite{hirsch_extended_1962} and Hirth \& Lothe~\cite{hirth_glide_1967} presented a systematic analysis of the dissociated jog structures in FCC metals based on the continuum theory of dislocations.
This analysis predicts two kinds of jogs, depending on the orientation relationship between the glide planes of the jog and the dislocation.
The ``acute'' jogs are more extended and glissile, while the ``obtuse'' jogs are more constricted and difficult to move.
%
%
%
Molecular dynamics (MD) simulations have largely confirmed these predictions, demonstrating a more pronounced drag effect from the ``obtuse'' jogs~\cite{rodney_dislocation_2000, abu-odeh_insights_2020, fey_accelerated_2021}.
Nonetheless, when moving, individual jogs on edge dislocations are still expected to glide conservatively with the edge dislocation~\cite{rodney_dislocation_2000}, involving no climb mechanisms nor vacancy emission.

In this work, we perform MD simulations of jogged edge dislocations in FCC nickel under different applied stress at room temperature ($\SI{300}{K}$) to study its mobility.
%
%
The applied stress is designed to produce solely an average glide force on the dislocation, without introducing any average climb force.
%
%
While our findings confirm the previous reports of jog behavior at low stresses, we discovered that at higher stress (e.g., $\qty{300}{MPa}$), the constricted jog not only glides but also intermittently climbs and emits vacancies, even at low temperatures ($\qty{300}{K}$).
%
%
%
%
%
%
%
We further demonstrated that the critical stress for the unexpected climb motion decreases with increasing the dislocation length, suggesting its potential relevance at even lower stresses in experimental samples.
%
%
Our analysis of the atomistic mechanisms underlying this climb process revealed it to be a stress-driven thermally activated process that competes with the glide motion.
These findings highlight the importance of dislocation-vacancy interactions, even at room temperature, in understanding plastic deformation, which was previously thought to be dominated solely by dislocation-dislocation interactions.
%

We use the LAMMPS simulation package~\cite{thompson_lammps_2022} with the embedded-atom method (EAM) interatomic potential developed by Angelo \emph{et al.}~\cite{angelo_trapping_1995}.
%
%
%
%
The simulation cell is oriented with the edge vectors in the $x$, $y$, $z$ directions aligned with $96[1\bar{1}0] \times 24[111] \times 84[\bar{1}\bar{1}2]$ in crystallographic directions, respectively.
The dislocation line and the Burgers vector are along the $z$- and $x$-axes, respectively, resulting in a slip plane normal to the $y$-axis.
Periodic boundary conditions are applied along $x$- and $z$- directions, while two free surfaces are introduced along the $y$-direction to accommodate a single dislocation.
The dislocation length is $L_z = \qty{36}{nm}$, and contains two unit-height jogs separated by $L_z / 2$.
%

%
%
%
%
\begin{figure}[!ht]
    \centering
    \includegraphics[width=0.6\linewidth]{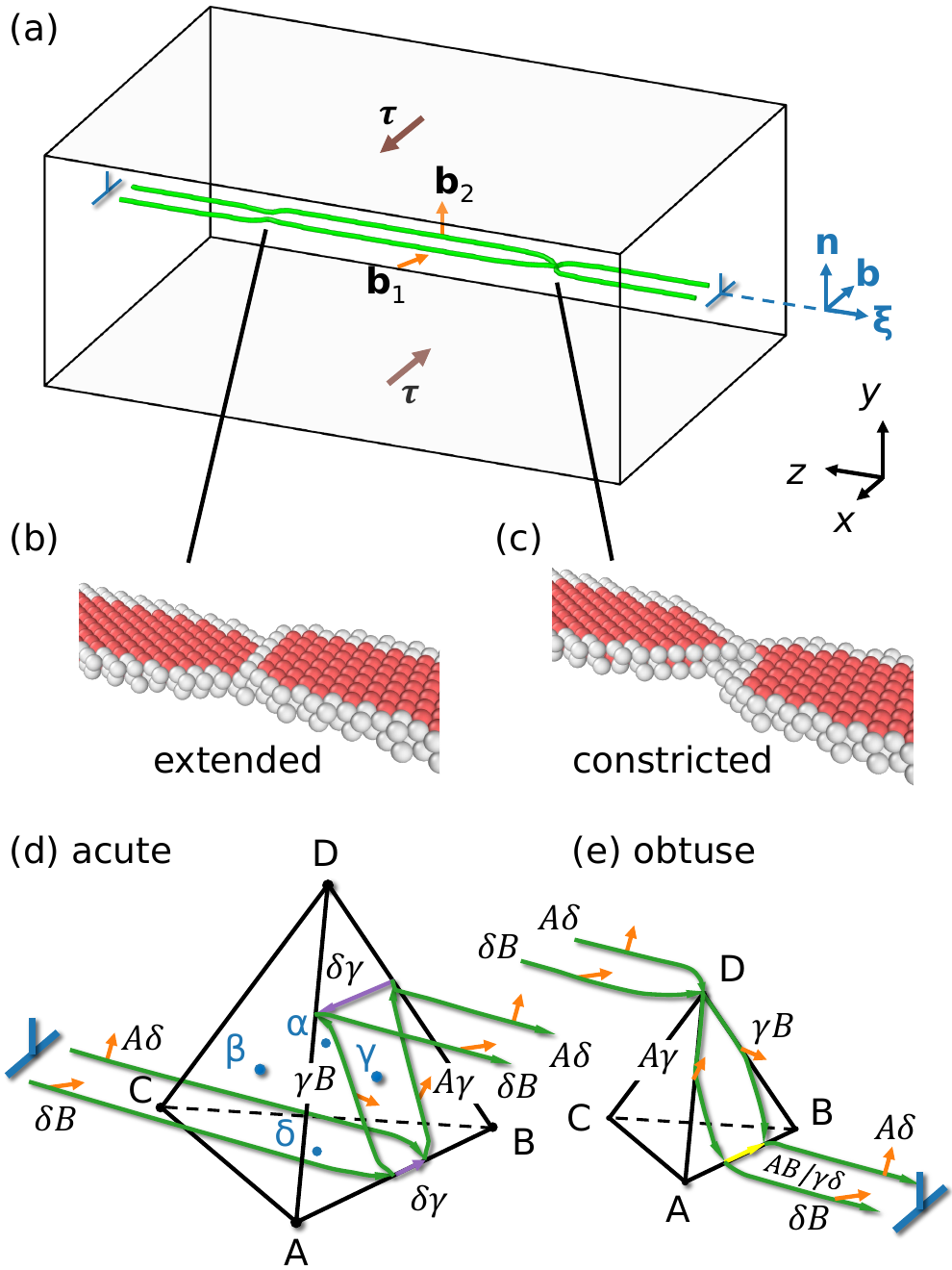}
    \caption{
        (a) Relaxed edge dislocation configuration with a unit jog pair.
        The Burgers vector $\mathbf{b}$ and the line vector $\bm{\xi}$ follows the right-hand, start-finish (RHSF) convention~\cite{cai_imperfections_2016}.
        The dislocation segments on the slip-plane dissociate into Shockley partials (orange arrows) $\mathbf{b}_1$ and $\mathbf{b}_2$.
        One of the jogs dissociates into an extended configuration while the other remains constricted.
        (b) Atomic structure of the extended jog.
        (c) Atomic structure of the constricted jog.
        The atoms are colored by their types from the common neighbor analysis.
        The face-centered cubic atoms (12 neighbors) are removed from the view to reveal the dislocation jog structures and stacking faults.
        %
        %
        %
        (d) Partial dislocation analysis of the extended (acute) jog.
        %
        %
        (e) Partial dislocation analysis of the constricted (obtuse) jog.
        %
        %
        %
        The green, yellow, and purple dislocation lines indicate the Shockley, Hirth, and stair-rod types of partial dislocations, respectively.
    }
    \label{fig:extended_jog}
\end{figure}

After introducing the jogged edge dislocation to the crystal, the atomic positions were first relaxed to an energy minimum.
Fig.~\ref{fig:extended_jog} shows the relaxed dislocation configuration.
The dislocation lines are extracted using the dislocation analysis (DXA) algorithm by OVITO~\cite{stukowski_visualization_2009, stukowski_automated_2012}.
The perfect edge
dislocation dissociates into two Shockley partials with Burgers vectors $\mathbf{b}_1$ = $\frac{1}{6}[\bar{1}2\bar{1}]$ and $\mathbf{b}_2$ = $\frac{1}{6}[\bar{2}11]$, bounding a stacking fault area.
%
%
Fig.~\ref{fig:extended_jog}(b) and (c) show the atomic structure of the two jogs, where one appears more extended than the other,
%
%
%
%
%
consistent with previous atomistic simulations~\cite{rodney_dislocation_2000, abu-odeh_insights_2020}.

We employ the continuum dislocation analysis with the Thompson's tetrahedron following the approach of Hirth and Lothe~\cite{hirth_theory_1982}.
Fig.~\ref{fig:extended_jog}(d) and (e) illustrate the jog configurations,
with most of the edge dislocation spreading on the $ABC$ plane.
Although the jog is only one atomic layer in height (unit jog), we assume that it lies on the $ACD$ plane where it dissociates into partials, consistent with previous simulation~\cite{rodney_dislocation_2000} and analysis~\cite{hirth_glide_1967} of superjog configurations.
The extended jog corresponds to the configuration shown in Fig.~\ref{fig:extended_jog}(d) with acute angles between the dislocation slip planes and is also called the acute jog.
The constricted jog corresponds to the configuration shown in Fig.~\ref{fig:extended_jog}(e) with obtuse angles between the dislocation slip planes and is also called the obtuse jog.
The obtuse jog is more constricted than the acute jog because of a larger Burgers vector and higher line energy for the Hirth partial at the intersection of the slip planes compared to the stair-rod in the acute jog:
$\|AB/\gamma\delta\|= \frac{\sqrt{2}}{3}b$ in the obtuse jog vs $\|\delta\gamma\| = \frac{1}{3}b$ in the acute jog, where $b=\frac{\sqrt{2}}{2}a$ is the magnitude of the Burgers vector of the perfect edge dislocation~\cite{hirsch_extended_1962}.
The upper and lower ends of the jogs also exhibit different dissociation widths due to the tendency of the Shockley partial dislocations on the slip planes ($ABC$ or $ABD$) to rotate toward screw orientations to minimize their line tension energies.
%
Unlike previous analyses, we find that the top part of the obtuse jog is fully constricted down to a single atom, as shown in Fig.~\ref{fig:extended_jog}(c).
As a result, the partials $A\gamma$ and $\gamma B$ join at the top of the jog and form a junction point (D).
As suggested by the Peierls-Nabarro model~\cite{bulatov_computer_2006}, a more widely dissociated dislocation is generally more mobile. 
Although both the extended (acute) and the constricted (obtuse) jogs are glissile on the $ACD$ slip plane, the constricted jog is expected to be more difficult to move (i.e., have a higher Peierls' barrier).
%

%
%
%

%
%
%
%
%
%
%


\begin{figure}[!ht]
    \centering
    \includegraphics[width=0.7\linewidth]{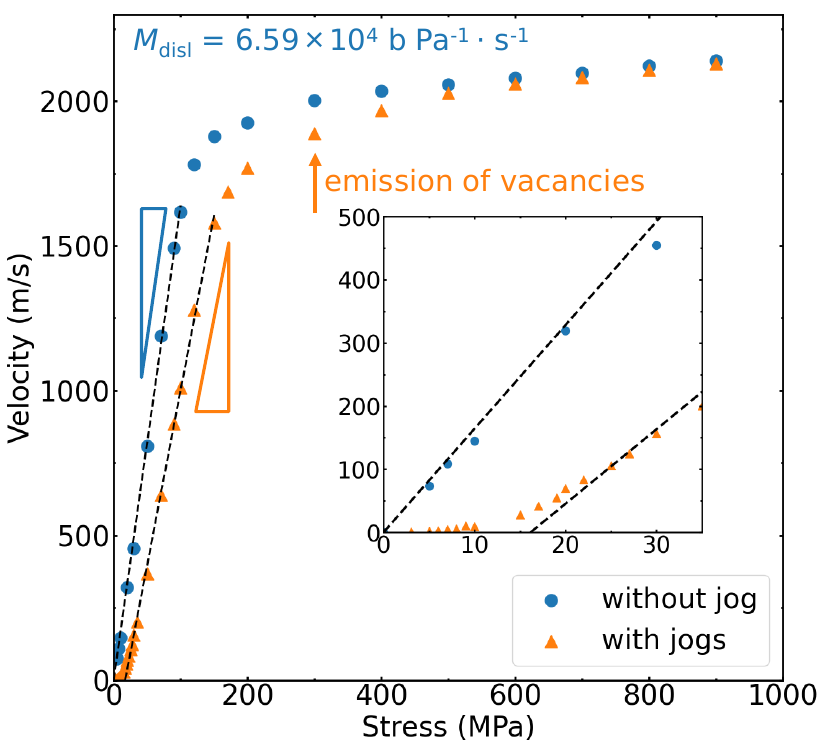}
    \caption{
    Dislocation velocity as a function of applied shear stress for edge dislocation (of length $\qty{36}{nm}$) with and without jogs.
    The arrow indicates the stress at and above which emission of vacancies from the constricted jog is observed.
    %
    %
    %
    %
    }
    \label{fig:disl_mobility}
\end{figure}

\begin{figure}[!ht]
    \centering
    \includegraphics[width=0.9\linewidth]{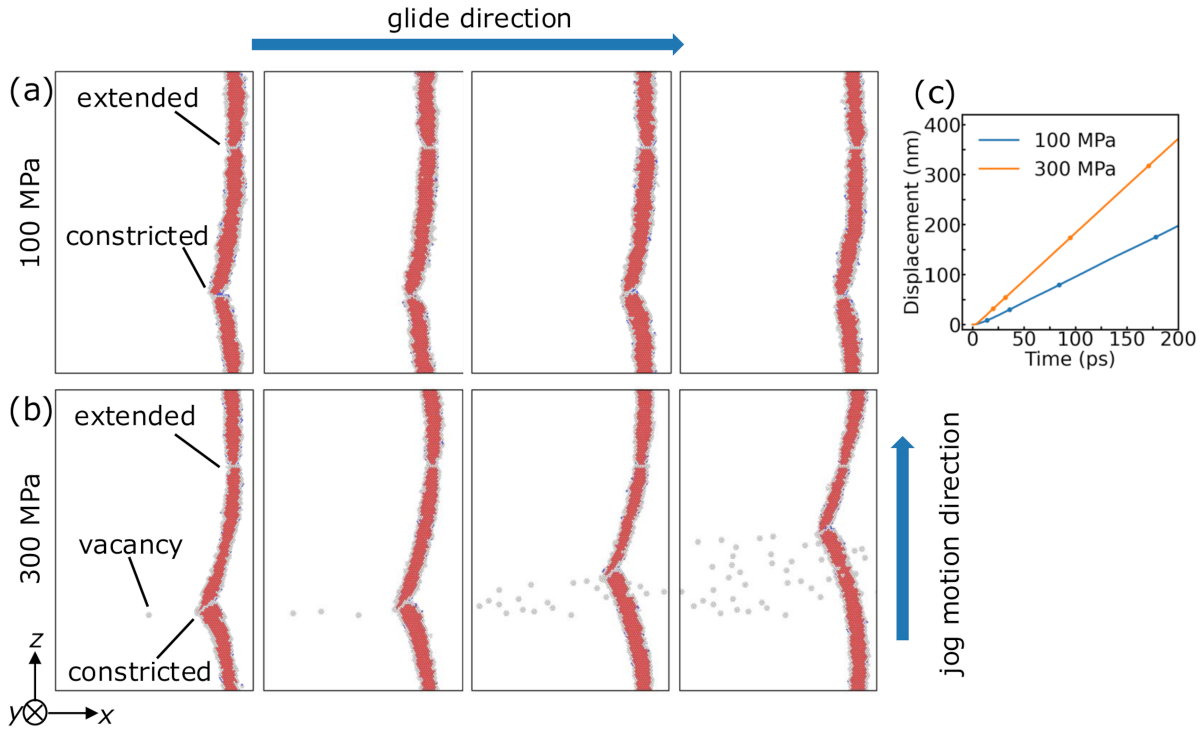}
    \caption{
    Atomic configurations during the motion of the jogged edge dislocation at the applied stresses of
    (a) $\SI{100}{MPa}$ and (b) $\SI{300}{MPa}$.
    %
    Atoms are colored in the same scheme as that
    in Fig.~\ref{fig:extended_jog}(b)(c).
    %
    (c) The displacement-time curves of the jogged edge dislocation at two different applied stresses.
    The dots correspond to the time at which the snapshots in (a) and (b) are taken.
    }
    \label{fig:disl_jog_configs}
\end{figure}

The relaxed dislocation configuration is then heated up to $\qty{300}{K}$ using NVT ensemble with the box size iteratively adjusted to reach zero stress for $\qty{300}{ps}$ until equilibrium.
To drive dislocation motion, a shear stress $\tau$ of different values is applied to the two surfaces along the Burgers vector direction by applying an external force to the surface atoms, as sketched in Fig.~\ref{fig:extended_jog}(a). 
%
%
%
%
%
%
%
At each applied stress value $\tau$, the dislocation velocity is obtained from the slope of the dislocation position over the time curve when the dislocation motion has reached a steady state, see Fig.~\ref{fig:disl_jog_configs}(c).
\correct{The simulation time goes up to $\qty{800}{ps}$ for low-stress conditions to ensure that steady-state motion is achieved (see Supplementary Figure~3).}
Fig.~\ref{fig:disl_mobility} plots the dislocation velocity as a function of applied stress for the jogged edge dislocation (orange triangles).
The velocities for a straight-edge dislocation (without jogs, blue dots) are also presented as a comparison.
At the low-stress regime (see the inset of Fig.~\ref{fig:disl_mobility}), while the velocity of a straight dislocation is proportional to the applied stress, the jogged dislocation demonstrates a non-linear velocity-stress relationship below $\qty{25}{MPa}$.
%
%
%
%
%
At intermediate stresses above $\qty{25}{MPa}$, the jogged dislocation also presents linear velocity-stress relationship, but at a lower slope 
than the straight dislocation ($\qty{6.59e4}{b\; Pa^{-1}\cdot s^{-1}}$).
At higher stresses (e.g., above $\qty{200}{MPa}$), the velocity-stress behavior of the jogged dislocation becomes qualitatively similar to that of the straight dislocation, where the linear region is replaced by a non-linear regime due to the relativistic effect, plateau towards the same limiting velocity of a straight edge dislocation~\cite{blaschke_how_2021}.
%
%
%
%

Fig.~\ref{fig:disl_jog_configs}(a) shows a few snapshots of the jogged dislocation configuration during motion at $\tau = \qty{100}{MPa}$, revealing significant bowing only near the constricted jog.
This indicates that only the constricted (obtuse) jog produces a significant drag on the jogged edge dislocation, while the extended jog appears to have minimal influence on the dislocation mobility.

\begin{figure}[!ht]
    \centering
    \includegraphics[width=0.8\linewidth]{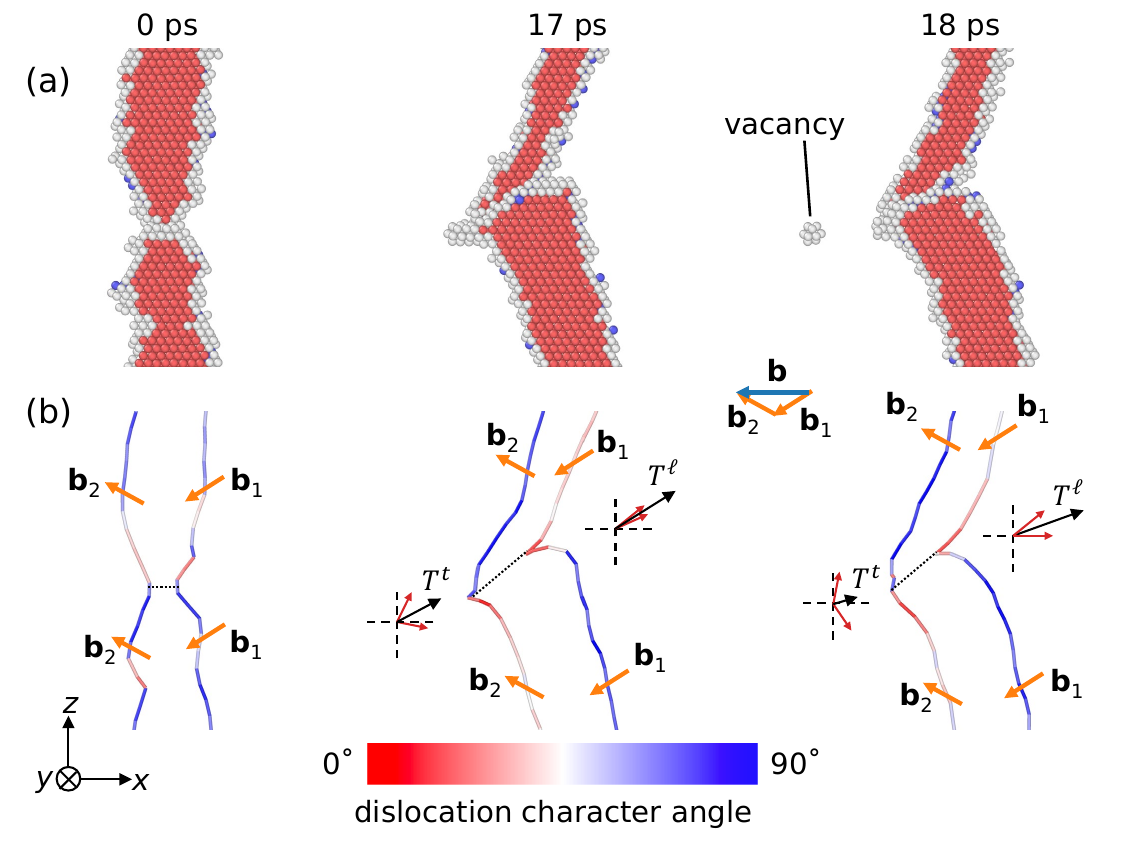}
    \caption{
    The first vacancy emission from edge jogged dislocation (at $0$, $17$, and $\qty{18}{ps}$) at the applied stress of $\qty{300}{MPa}$:
    (a) atomic configurations and 
    (b) dislocation line structures colored by their characteristic angle.
    %
    \correct{The line tension forces from the leading and trailing partials are noted as $T^{\ell}$ and $T^{t}$, respectively.}
    %
    %
    \correct{The black dotted line connecting the two partials represents the jog location between the two stacking faults on the two slip planes.}
    }
    \label{fig:disl_vacancies}
\end{figure}
Fig.~\ref{fig:disl_jog_configs}(b) shows snapshots of the jogged dislocation configuration during motion at $\tau = \qty{300}{MPa}$.
Unexpectedly, in the first frame shown here, the constricted jog climbed along the dislocation line and emitted a vacancy, after it had glided forward conservatively for around $\qty{15}{ps}$.  The jog climbs further along the dislocation line ($z$-axis) and emits more vacancies as it glides further, as shown in the subsequent frames.
At higher stresses, the constricted jog's climb motion and vacancy emission become more frequent,
implying a stress-driven thermally-activated mechanism.
%
%
Although the critical stress for the climb motion and vacancy emission is about $\qty{300}{MPa}$, slightly higher than the linear velocity-stress regime relevant to crystal plasticity under typical loading conditions, it may be attributed to the short dislocation length ($L_z = \qty{36}{nm}$) 
in the simulation cell.
By repeating the MD simulations of jogged dislocation with a longer dislocation length of $L_z = \qty{72}{nm}$,
%
we observe \correct{a lower critical stress of dislocation climb and vacancy emission at $\qty{200}{MPa}$.}
\correct{The origin of this length dependence will be discussed below.  Here we give a rough estimate of the critical stress in experimental settings, where dislocation lines are typically several hundreds of nanometers to micrometers long.}
%
%
%
%
\correct{
Assuming the critical stress to follow a power law with the dislocation length, i.e., $\sigma_c \propto l^{-\alpha}$, then extrapolating our data points to $l = \qty{1}{\micro\meter}$ gives an estimated critical stress of $\qty{44}{MPa}$,
%
%
%
which is relevant to the deformation experiments under ambient conditions.
}
%

\correct{To rule out the possibility of artifacts from the interatomic potential used, 
we re-run the MD simulation at $T = \qty{300}{K}$ and $\tau = \qty{300}{MPa}$ using another EAM potential~\cite{rao_atomistic_1999} that has been used in recent dislocation simulations~\cite{wang_stress-dependent_2023}.
%
%
The constricted jog exhibits the same vacancy emission behavior, as shown in Supplementary Figure~5.
}


In the following, we explain the underlying mechanism for
the unexpected vacancy emission of the jogged edge dislocation.
For the constricted jog at small stresses, due to its higher Peierls barrier,
the conservative glide motion becomes a thermally activated process.
%
%
The applied (Schmid) stress is expected to lower the Peierls barrier of the jog's glide motion.
At around \correct{$\qty{25}{MPa}$} stress the energy barrier for the glide motion of the jog has become negligible.
%
%
At the intermediate stress with a linear velocity-stress relationship,
the dislocation velocity is limited by the viscous dragging effects from finite-temperature effects through interactions with phonons~\cite{hirth_theory_1982}.
At these stresses, even though the Peierls barrier for the constricted jog has been overcome by stress, the constricted jog still experiences higher drag than the rest of the edge dislocation.
As a result, the jogged dislocation moves consistently slower than the straight dislocation at the same stress until about $\qty{600}{MPa}$ where the relativistic effect dominates.

Above the critical stress where the constricted jog climbs and emits vacancies,
it appears that this process is thermally activated and hence stochastic, as the constricted jog can glide for some distance without climbing.
It seems counterintuitive that the applied (Schmid) stress, which is supposed to drive the glide motion, also enhances the rate of climb motion by lowering its energy barrier,
%
%
even though the applied stress only generates a glide Peach-Koehler force on the jogged edge dislocation.

%
\correct{The driving force for the climb motion of the dislocation jog can be intuitively understood from a simple line-tension model.
}
Fig.~\ref{fig:disl_vacancies} shows the atomic configuration and the corresponding dislocation structures of the constricted jog during the MD simulation at an applied stress of $\qty{300}{MPa}$.
As the dislocation moves, due to the slower velocity of the constricted jog, the dislocation line bows locally near the constricted jog.
%
%
Due to the dissociation of the perfect dislocation into Shockley partials, and the preference for dislocations to adopt the screw orientation (i.e. parallel to its own Burgers vector), the bowing of the dislocation lines on the two sides of the jog is asymmetrical.
%
%
%
%
As shown in Fig.~\ref{fig:disl_vacancies}(b), the bowing of the partials on the $-z$ side of the jog is more severe than those on the \correct{$+z$} side according to the dislocation analysis from OVITO~\cite{stukowski_visualization_2009, stukowski_automated_2012}h.
%
%
%
\correct{For the leading partial, both arms become almost parallel to its Burgers vector $\mathbf{b}_1$, resulting in a strong line-tension force $T^{\ell}$ on the jog with a large $+z$ component.}
\correct{For the trailing partial, the dislocation line on the $+z$ side of the jog rotates to a near-edge orientation while the other side becomes nearly screw (i.e. parallel to $\mathbf{b}_2$); the resulting line tension force $T^{\rm t}$ still has a non-zero $+z$ component.}
%
%
\correct{Therefore, the line tensions from both leading and trailing partials result in forces pulling the jog in the $+z$ direction, which is a climb direction for the jog.}
%
%
%
\correct{We hypothesize that vacancy emission occurs when the dislocation lines next to the jog have rotated to the critical orientations described above. Since a longer dislocation line can achieve the same bowing angle at a lower stress (like a Frank-Read source), the critical stress for vacancy emission is expected to decrease with increasing dislocation length between the two constricted jogs.}
\correct{Therefore, we believe the climb motion of the jog is due to the line-tension effect from the asymmetric bowing of the dislocation lines, and that these line tension effects are much stronger than long-range elastic interactions between jogs and their periodic images~\cite{ji_quantifying_2020}.}
\correct{More work is needed to test this hypothesis and obtain a quantitative understanding of the vacancy emission mechanism, such as by extracting the effective local stress component~\cite{ji_atomistic--microscale_2024} that induces climb motion.}

\correct{In the MD simulations reported here,
although the dislocation moves cross the periodic boundaries and returns to the location where it previously emitted the vacancies,
there is no pronounced dislocation-vacancy interaction observed at $\qty{300}{K}$ and $\qty{300}{MPa}$, as shown in {Supplementary Figure~4(a)}.
The vacancies happen to miss the dislocation in their subsequent encounters.
}
\correct{
The vacancy-dislocation interaction is not observed except for very high-stress simulation cases (e.g., $\qty{900}{MPa}$) and at a late state of the simulation, see Supplementary Figure~4(b).
When the constricted jog has climbed so much and gets very close to the extended jog, the jog pair annihilates and forms a vacancy cluster.
This vacancy cluster (or part of it) can be re-absorbed by the dislocation when the dislocation moves cross the periodic boundary.
We have excluded such configurations from the calculation of dislocation velocities (see {Supplementary Materials, Section~2}).
}


\correct{
Experiments have shown vacancy generation during room-temperature plastic deformation using a variety of techniques~\cite{cizek_development_2019, ungar_vacancy_2007, zehetbauer_cold_1993}.}
\correct{However, it was commonly thought that only jogs on screw dislocations may produce vacancies by climb.  Among dislocations with all character angles, edge dislocations were considered least likely to produce vacancies from jogs.  Our finding that even edge dislocations can produce vacancies from jogs implies that all dislocations (including mixed dislocations) can produce vacancies from jogs.  This is qualitative consistent with the experimental finding~\cite{cizek_development_2019} that vacancy production is common in FCC metals, whose microstructure is not dominated by screw dislocations.}

In summary, we observed vacancy emission from a jog on an edge dislocation in an FCC metal from MD simulations at $\qty{300}{K}$.
The physical mechanism is ultimately traced to the asymmetric bowing of the partial dislocations around the constricted jog which drags the moving edge dislocation.
This finding reveals a previously unexpected role of point defects in the work hardening of FCC metals under ambient conditions.
In addition, the vacancies emitted by the jogs can trap dissolved hydrogen atoms to form embryos of microvoids, playing a role in hydrogen-embrittlement of FCC metals~\cite{lu_hydrogen_2005}.
%
%
The results highlight the need for a more systematic study of the jog effects on dislocation motion on all types of dislocations in FCC metals.
%


\textbf{Acknowledgements}

This work was supported by the National Science Foundation under Award Number DMREF 2118522 (W.J. and W.C.).
Y. W. acknowledges the support from 
the Stanford Energy Postdoctoral Fellowship and the Precourt Institute for Energy.

\bibliographystyle{elsarticle-num-names}
\bibliography{jog.bib}

\begin{thebibliography}{38}
\expandafter\ifx\csname natexlab\endcsname\relax\def\natexlab#1{#1}\fi
\providecommand{\url}[1]{\texttt{#1}}
\providecommand{\href}[2]{#2}
\providecommand{\path}[1]{#1}
\providecommand{\DOIprefix}{doi:}
\providecommand{\ArXivprefix}{arXiv:}
\providecommand{\URLprefix}{URL: }
\providecommand{\Pubmedprefix}{pmid:}
\providecommand{\doi}[1]{\href{http://dx.doi.org/#1}{\path{#1}}}
\providecommand{\Pubmed}[1]{\href{pmid:#1}{\path{#1}}}
\providecommand{\bibinfo}[2]{#2}
\ifx\xfnm\relax \def\xfnm[#1]{\unskip,\space#1}\fi
\bibitem[{Cai and Nix(2016)}]{cai_imperfections_2016}
\bibinfo{author}{W.~Cai}, \bibinfo{author}{W.~D. Nix},
  \bibinfo{title}{Imperfections in {Crystalline} {Solids}},
  \bibinfo{publisher}{Cambridge University Press}, \bibinfo{year}{2016}.
\bibitem[{Hull and Bacon(2011)}]{hull_introduction_2011}
\bibinfo{author}{D.~Hull}, \bibinfo{author}{D.~J. Bacon},
  \bibinfo{title}{Introduction to dislocations}, \bibinfo{edition}{5th} ed.,
  \bibinfo{publisher}{Butterworth Heinemann, Elsevier},
  \bibinfo{address}{Amsterdam Heidelberg}, \bibinfo{year}{2011}.
\bibitem[{Messerschmidt(1970)}]{messerschmidt_model_1970}
\bibinfo{author}{U.~Messerschmidt},
\newblock \bibinfo{title}{A {Model} of the {Temperature} {Dependent} {Part} of
  {Stage} {I} {Work}-{Hardening} due to {Jog}-{Dragging} ({I})},
\newblock \bibinfo{journal}{physica status solidi (b)} \bibinfo{volume}{41}
  (\bibinfo{year}{1970}) \bibinfo{pages}{549--563}.
\bibitem[{Abu-Odeh et~al.(2020)Abu-Odeh, Cottura, and
  Asta}]{abu-odeh_insights_2020}
\bibinfo{author}{A.~Abu-Odeh}, \bibinfo{author}{M.~Cottura},
  \bibinfo{author}{M.~Asta},
\newblock \bibinfo{title}{Insights into dislocation climb efficiency in {FCC}
  metals from atomistic simulations},
\newblock \bibinfo{journal}{Acta Materialia} \bibinfo{volume}{193}
  (\bibinfo{year}{2020}) \bibinfo{pages}{172--181}.
\bibitem[{Vegge et~al.(2001)Vegge, Rasmussen, Leffers, Pedersen, and
  Jacobsen}]{vegge_atomistic_2001}
\bibinfo{author}{T.~Vegge}, \bibinfo{author}{T.~Rasmussen},
  \bibinfo{author}{T.~Leffers}, \bibinfo{author}{O.~B. Pedersen},
  \bibinfo{author}{K.~W. Jacobsen},
\newblock \bibinfo{title}{Atomistic simulations of cross-slip of jogged screw
  dislocations in copper},
\newblock \bibinfo{journal}{Philosophical Magazine Letters}
  \bibinfo{volume}{81} (\bibinfo{year}{2001}) \bibinfo{pages}{137--144}.
\bibitem[{Barrett and Nix(1965)}]{barrett_model_1965}
\bibinfo{author}{C.~R. Barrett}, \bibinfo{author}{W.~D. Nix},
\newblock \bibinfo{title}{A model for steady state creep based on the motion of
  jogged screw dislocations},
\newblock \bibinfo{journal}{Acta Metallurgica} \bibinfo{volume}{13}
  (\bibinfo{year}{1965}) \bibinfo{pages}{1247--1258}.
\bibitem[{Zbib et~al.(2000)Zbib, Diaz de~la Rubia, Rhee, and
  P.~Hirth}]{zbib_3d_2000}
\bibinfo{author}{H.~M. Zbib}, \bibinfo{author}{T.~Diaz de~la Rubia},
  \bibinfo{author}{M.~Rhee}, \bibinfo{author}{J.~P.~Hirth},
\newblock \bibinfo{title}{{3D} dislocation dynamics: stress{\textendash}strain
  behavior and hardening mechanisms in fcc and bcc metals},
\newblock \bibinfo{journal}{Journal of Nuclear Materials} \bibinfo{volume}{276}
  (\bibinfo{year}{2000}) \bibinfo{pages}{154--165}.
\bibitem[{Rodney and Martin(1999)}]{rodney_dislocation_1999}
\bibinfo{author}{D.~Rodney}, \bibinfo{author}{G.~Martin},
\newblock \bibinfo{title}{Dislocation {Pinning} by {Small} {Interstitial}
  {Loops}: {A} {Molecular} {Dynamics} {Study}},
\newblock \bibinfo{journal}{Physical Review Letters} \bibinfo{volume}{82}
  (\bibinfo{year}{1999}) \bibinfo{pages}{3272--3275}.
\bibitem[{Bertin et~al.(2020)Bertin, Sills, and Cai}]{bertin_frontiers_2020}
\bibinfo{author}{N.~Bertin}, \bibinfo{author}{R.~B. Sills},
  \bibinfo{author}{W.~Cai},
\newblock \bibinfo{title}{Frontiers in the simulation of dislocations},
\newblock \bibinfo{journal}{Annual Review of Materials Research}
  \bibinfo{volume}{50} (\bibinfo{year}{2020}) \bibinfo{pages}{437--464}.
\bibitem[{Po et~al.(2022)Po, Huang, Li, Baker, Flores, Black, Hollenbeck, and
  Ghoniem}]{po_model_2022}
\bibinfo{author}{G.~Po}, \bibinfo{author}{Y.~Huang}, \bibinfo{author}{Y.~Li},
  \bibinfo{author}{K.~Baker}, \bibinfo{author}{B.~R. Flores},
  \bibinfo{author}{T.~Black}, \bibinfo{author}{J.~Hollenbeck},
  \bibinfo{author}{N.~Ghoniem},
\newblock \bibinfo{title}{A model of thermal creep and annealing in finite
  domains based on coupled dislocation climb and vacancy diffusion},
\newblock \bibinfo{journal}{Journal of the Mechanics and Physics of Solids}
  \bibinfo{volume}{169} (\bibinfo{year}{2022}) \bibinfo{pages}{105066}.
\bibitem[{Li et~al.(2023)Li, Ghoniem, Baker, Ramirez~Flores, Black, Hollenbeck,
  and Po}]{li_coupled_2023}
\bibinfo{author}{Y.~Li}, \bibinfo{author}{N.~Ghoniem},
  \bibinfo{author}{K.~Baker}, \bibinfo{author}{B.~Ramirez~Flores},
  \bibinfo{author}{T.~Black}, \bibinfo{author}{J.~Hollenbeck},
  \bibinfo{author}{G.~Po},
\newblock \bibinfo{title}{A coupled vacancy diffusion-dislocation dynamics
  model for the climb-glide motion of jogged screw dislocations},
\newblock \bibinfo{journal}{Acta Materialia} \bibinfo{volume}{244}
  (\bibinfo{year}{2023}) \bibinfo{pages}{118546}.
\bibitem[{Cai et~al.(2004)Cai, Bulatov, Chang, Li, and Yip}]{cai_chapter_2004}
\bibinfo{author}{W.~Cai}, \bibinfo{author}{V.~V. Bulatov},
  \bibinfo{author}{J.~Chang}, \bibinfo{author}{J.~Li},
  \bibinfo{author}{S.~Yip},
\newblock \bibinfo{title}{Chapter 64 - {Dislocation} {Core} {Effects} on
  {Mobility}},
\newblock in: \bibinfo{editor}{F.~R.~N. Nabarro}, \bibinfo{editor}{J.~P. Hirth}
  (Eds.), \bibinfo{booktitle}{Dislocations in {Solids}},
  volume~\bibinfo{volume}{12} of \textit{\bibinfo{series}{Dislocations in
  {Solids}}}, \bibinfo{publisher}{Elsevier}, \bibinfo{year}{2004}, pp.
  \bibinfo{pages}{1--80}.
\bibitem[{Gu et~al.(2015)Gu, Xiang, Quek, and
  Srolovitz}]{gu_three-dimensional_2015}
\bibinfo{author}{Y.~Gu}, \bibinfo{author}{Y.~Xiang}, \bibinfo{author}{S.~S.
  Quek}, \bibinfo{author}{D.~J. Srolovitz},
\newblock \bibinfo{title}{Three-dimensional formulation of dislocation climb},
\newblock \bibinfo{journal}{Journal of the Mechanics and Physics of Solids}
  \bibinfo{volume}{83} (\bibinfo{year}{2015}) \bibinfo{pages}{319--337}.
\bibitem[{Breidi and Dudarev(2022)}]{breidi_dislocation_2022}
\bibinfo{author}{A.~Breidi}, \bibinfo{author}{S.~L. Dudarev},
\newblock \bibinfo{title}{Dislocation dynamics simulation of thermal annealing
  of a dislocation loop microstructure},
\newblock \bibinfo{journal}{Journal of Nuclear Materials} \bibinfo{volume}{562}
  (\bibinfo{year}{2022}) \bibinfo{pages}{153552}.
\bibitem[{Hirth and Lothe(1967)}]{hirth_glide_1967}
\bibinfo{author}{J.~P. Hirth}, \bibinfo{author}{J.~Lothe},
\newblock \bibinfo{title}{Glide of jogged dislocations},
\newblock \bibinfo{journal}{Canadian Journal of Physics} \bibinfo{volume}{45}
  (\bibinfo{year}{1967}) \bibinfo{pages}{809--826}.
\bibitem[{Strunk and Frydman(1975)}]{strunk_jog_1975}
\bibinfo{author}{H.~Strunk}, \bibinfo{author}{R.~Frydman},
\newblock \bibinfo{title}{Jog dragging in edge dislocations with application to
  plastic deformation and internal friction},
\newblock \bibinfo{journal}{Materials Science and Engineering}
  \bibinfo{volume}{18} (\bibinfo{year}{1975}) \bibinfo{pages}{143--148}.
\bibitem[{Seeger(1955)}]{seeger_cxxxii_1955}
\bibinfo{author}{A.~Seeger},
\newblock \bibinfo{title}{{CXXXII}. {The} generation of lattice defects by
  moving dislocations, and its application to the temperature dependence of the
  flow-stress of {F}.{C}.{C}. crystals},
\newblock \bibinfo{journal}{The London, Edinburgh, and Dublin Philosophical
  Magazine and Journal of Science} \bibinfo{volume}{46} (\bibinfo{year}{1955})
  \bibinfo{pages}{1194--1217}.
\bibitem[{Akhondzadeh et~al.(2023)Akhondzadeh, Kang, Sills, Ramesh, and
  Cai}]{akhondzadeh_direct_2023}
\bibinfo{author}{S.~Akhondzadeh}, \bibinfo{author}{M.~Kang},
  \bibinfo{author}{R.~B. Sills}, \bibinfo{author}{K.~T. Ramesh},
  \bibinfo{author}{W.~Cai},
\newblock \bibinfo{title}{Direct comparison between experiments and dislocation
  dynamics simulations of high rate deformation of single crystal copper},
\newblock \bibinfo{journal}{Acta Materialia} \bibinfo{volume}{250}
  (\bibinfo{year}{2023}) \bibinfo{pages}{118851}.
\bibitem[{Messerschmidt(1971)}]{messerschmidt_model_1971}
\bibinfo{author}{U.~Messerschmidt},
\newblock \bibinfo{title}{A model of the temperature dependent part of stage
  {I} work-hardening due to jog-dragging ({II})},
\newblock \bibinfo{journal}{physica status solidi (b)} \bibinfo{volume}{48}
  (\bibinfo{year}{1971}) \bibinfo{pages}{781--790}.
\bibitem[{Carter(1979)}]{carter_extension_1979}
\bibinfo{author}{C.~B. Carter},
\newblock \bibinfo{title}{The {Extension} of {Jogs} on {Dissociated}
  {Dislocations} in {F}.{C}.{C}. {Metals}},
\newblock \bibinfo{journal}{physica status solidi (a)} \bibinfo{volume}{54}
  (\bibinfo{year}{1979}) \bibinfo{pages}{395--406}.
\bibitem[{Hirsch(1962)}]{hirsch_extended_1962}
\bibinfo{author}{P.~B. Hirsch},
\newblock \bibinfo{title}{Extended jogs in dislocations in face-centred cubic
  metals},
\newblock \bibinfo{journal}{The Philosophical Magazine: A Journal of
  Theoretical Experimental and Applied Physics} \bibinfo{volume}{7}
  (\bibinfo{year}{1962}) \bibinfo{pages}{67--93}.
\bibitem[{Rodney and Martin(2000)}]{rodney_dislocation_2000}
\bibinfo{author}{D.~Rodney}, \bibinfo{author}{G.~Martin},
\newblock \bibinfo{title}{Dislocation pinning by glissile interstitial loops in
  a nickel crystal: {A} molecular-dynamics study},
\newblock \bibinfo{journal}{Physical Review B} \bibinfo{volume}{61}
  (\bibinfo{year}{2000}) \bibinfo{pages}{8714--8725}.
\bibitem[{Fey et~al.(2021)Fey, Tan, Swinburne, Perez, and
  Trinkle}]{fey_accelerated_2021}
\bibinfo{author}{L.~T.~W. Fey}, \bibinfo{author}{A.~M.~Z. Tan},
  \bibinfo{author}{T.~D. Swinburne}, \bibinfo{author}{D.~Perez},
  \bibinfo{author}{D.~R. Trinkle},
\newblock \bibinfo{title}{Accelerated molecular dynamics simulations of
  dislocation climb in nickel},
\newblock \bibinfo{journal}{Physical Review Materials} \bibinfo{volume}{5}
  (\bibinfo{year}{2021}) \bibinfo{pages}{083603}.
\bibitem[{Thompson et~al.(2022)Thompson, Aktulga, Berger, Bolintineanu, Brown,
  Crozier, in~'t Veld, Kohlmeyer, Moore, Nguyen, Shan, Stevens, Tranchida,
  Trott, and Plimpton}]{thompson_lammps_2022}
\bibinfo{author}{A.~P. Thompson}, \bibinfo{author}{H.~M. Aktulga},
  \bibinfo{author}{R.~Berger}, \bibinfo{author}{D.~S. Bolintineanu},
  \bibinfo{author}{W.~M. Brown}, \bibinfo{author}{P.~S. Crozier},
  \bibinfo{author}{P.~J. in~'t Veld}, \bibinfo{author}{A.~Kohlmeyer},
  \bibinfo{author}{S.~G. Moore}, \bibinfo{author}{T.~D. Nguyen},
  \bibinfo{author}{R.~Shan}, \bibinfo{author}{M.~J. Stevens},
  \bibinfo{author}{J.~Tranchida}, \bibinfo{author}{C.~Trott},
  \bibinfo{author}{S.~J. Plimpton},
\newblock \bibinfo{title}{{LAMMPS} - a flexible simulation tool for
  particle-based materials modeling at the atomic, meso, and continuum scales},
\newblock \bibinfo{journal}{Computer Physics Communications}
  \bibinfo{volume}{271} (\bibinfo{year}{2022}) \bibinfo{pages}{108171}.
\bibitem[{Angelo et~al.(1995)Angelo, Moody, and Baskes}]{angelo_trapping_1995}
\bibinfo{author}{J.~E. Angelo}, \bibinfo{author}{N.~R. Moody},
  \bibinfo{author}{M.~I. Baskes},
\newblock \bibinfo{title}{Trapping of hydrogen to lattice defects in nickel},
\newblock \bibinfo{journal}{Modelling and Simulation in Materials Science and
  Engineering} \bibinfo{volume}{3} (\bibinfo{year}{1995}) \bibinfo{pages}{289}.
\bibitem[{Stukowski(2009)}]{stukowski_visualization_2009}
\bibinfo{author}{A.~Stukowski},
\newblock \bibinfo{title}{Visualization and analysis of atomistic simulation
  data with {OVITO}{\textendash}the {Open} {Visualization} {Tool}},
\newblock \bibinfo{journal}{Modelling and Simulation in Materials Science and
  Engineering} \bibinfo{volume}{18} (\bibinfo{year}{2009})
  \bibinfo{pages}{015012}.
\bibitem[{Stukowski et~al.(2012)Stukowski, Bulatov, and
  Arsenlis}]{stukowski_automated_2012}
\bibinfo{author}{A.~Stukowski}, \bibinfo{author}{V.~V. Bulatov},
  \bibinfo{author}{A.~Arsenlis},
\newblock \bibinfo{title}{Automated identification and indexing of dislocations
  in crystal interfaces},
\newblock \bibinfo{journal}{Modelling and Simulation in Materials Science and
  Engineering} \bibinfo{volume}{20} (\bibinfo{year}{2012})
  \bibinfo{pages}{085007}.
\bibitem[{Hirth and Lothe(1982)}]{hirth_theory_1982}
\bibinfo{author}{J.~P. Hirth}, \bibinfo{author}{J.~Lothe},
  \bibinfo{title}{Theory of dislocations}, \bibinfo{edition}{2nd} ed.,
  \bibinfo{publisher}{Wiley}, \bibinfo{address}{New York},
  \bibinfo{year}{1982}.
\bibitem[{Bulatov and Cai(2006)}]{bulatov_computer_2006}
\bibinfo{author}{V.~Bulatov}, \bibinfo{author}{W.~Cai},
  \bibinfo{title}{Computer {Simulations} of {Dislocations}},
  \bibinfo{publisher}{Oxford University Press}, \bibinfo{year}{2006}.
\bibitem[{Blaschke(2021)}]{blaschke_how_2021}
\bibinfo{author}{D.~N. Blaschke},
\newblock \bibinfo{title}{How to determine limiting velocities of dislocations
  in anisotropic crystals},
\newblock \bibinfo{journal}{Journal of Physics: Condensed Matter}
  \bibinfo{volume}{33} (\bibinfo{year}{2021}) \bibinfo{pages}{503005}.
\bibitem[{Rao et~al.(1999)Rao, Parthasarathy, and
  Woodward}]{rao_atomistic_1999}
\bibinfo{author}{S.~Rao}, \bibinfo{author}{T.~A. Parthasarathy},
  \bibinfo{author}{C.~Woodward},
\newblock \bibinfo{title}{Atomistic simulation of cross-slip processes in model
  fcc structures},
\newblock \bibinfo{journal}{Philosophical Magazine A} \bibinfo{volume}{79}
  (\bibinfo{year}{1999}) \bibinfo{pages}{1167--1192}.
\bibitem[{Wang and Cai(2023)}]{wang_stress-dependent_2023}
\bibinfo{author}{Y.~Wang}, \bibinfo{author}{W.~Cai},
\newblock \bibinfo{title}{Stress-dependent activation entropy in thermally
  activated cross-slip of dislocations},
\newblock \bibinfo{journal}{Proceedings of the National Academy of Sciences}
  \bibinfo{volume}{120} (\bibinfo{year}{2023}) \bibinfo{pages}{e2222039120}.
\bibitem[{Ji et~al.(2020)Ji, Phan, Chen, and Xiong}]{ji_quantifying_2020}
\bibinfo{author}{R.~Ji}, \bibinfo{author}{T.~Phan}, \bibinfo{author}{H.~Chen},
  \bibinfo{author}{L.~Xiong},
\newblock \bibinfo{title}{Quantifying the dynamics of dislocation kinks in iron
  and tungsten through atomistic simulations},
\newblock \bibinfo{journal}{International Journal of Plasticity}
  \bibinfo{volume}{128} (\bibinfo{year}{2020}) \bibinfo{pages}{102675}.
\bibitem[{Ji et~al.(2024)Ji, Phan, Chen, McDowell, and
  Xiong}]{ji_atomistic--microscale_2024}
\bibinfo{author}{R.~Ji}, \bibinfo{author}{T.~Phan}, \bibinfo{author}{Y.~Chen},
  \bibinfo{author}{D.~L. McDowell}, \bibinfo{author}{L.~Xiong},
\newblock \bibinfo{title}{An atomistic-to-microscale characterization of the
  kink-controlled dislocation dynamics in bcc metals through finite-temperature
  coarse-grained atomistic simulations},
\newblock \bibinfo{journal}{Acta Materialia} \bibinfo{volume}{262}
  (\bibinfo{year}{2024}) \bibinfo{pages}{119440}.
\bibitem[{{\v C}{\'i}{\v z}ek et~al.(2019){\v C}{\'i}{\v z}ek, Jane{\v c}ek,
  Vlas{\'a}k, Smola, Melikhova, Islamgaliev, and
  Dobatkin}]{cizek_development_2019}
\bibinfo{author}{J.~{\v C}{\'i}{\v z}ek}, \bibinfo{author}{M.~Jane{\v c}ek},
  \bibinfo{author}{T.~Vlas{\'a}k}, \bibinfo{author}{B.~Smola},
  \bibinfo{author}{O.~Melikhova}, \bibinfo{author}{R.~Islamgaliev},
  \bibinfo{author}{S.~Dobatkin},
\newblock \bibinfo{title}{The {Development} of {Vacancies} during {Severe}
  {Plastic} {Deformation}},
\newblock \bibinfo{journal}{Materials Transactions} \bibinfo{volume}{60}
  (\bibinfo{year}{2019}) \bibinfo{pages}{1533--1542}.
\bibitem[{Ung{\'a}r et~al.(2007)Ung{\'a}r, Schafler, Han{\'a}k, Bernstorff, and
  Zehetbauer}]{ungar_vacancy_2007}
\bibinfo{author}{T.~Ung{\'a}r}, \bibinfo{author}{E.~Schafler},
  \bibinfo{author}{P.~Han{\'a}k}, \bibinfo{author}{S.~Bernstorff},
  \bibinfo{author}{M.~Zehetbauer},
\newblock \bibinfo{title}{Vacancy production during plastic deformation in
  copper determined by in situ {X}-ray diffraction},
\newblock \bibinfo{journal}{Materials Science and Engineering: A}
  \bibinfo{volume}{462} (\bibinfo{year}{2007}) \bibinfo{pages}{398--401}.
\bibitem[{Zehetbauer(1993)}]{zehetbauer_cold_1993}
\bibinfo{author}{M.~Zehetbauer},
\newblock \bibinfo{title}{Cold work hardening in stages {IV} and {V} of
  {F}.{C}.{C}. metals{\textemdash}{II}. {Model} fits and physical results},
\newblock \bibinfo{journal}{Acta Metallurgica et Materialia}
  \bibinfo{volume}{41} (\bibinfo{year}{1993}) \bibinfo{pages}{589--599}.
\bibitem[{Lu and Kaxiras(2005)}]{lu_hydrogen_2005}
\bibinfo{author}{G.~Lu}, \bibinfo{author}{E.~Kaxiras},
\newblock \bibinfo{title}{Hydrogen {Embrittlement} of {Aluminum}: {The}
  {Crucial} {Role} of {Vacancies}},
\newblock \bibinfo{journal}{Physical Review Letters} \bibinfo{volume}{94}
  (\bibinfo{year}{2005}) \bibinfo{pages}{155501}.

\end{thebibliography}

\end{document}